\documentclass[12pt,preprint]{aastex}
\shorttitle{Vertical Structure of Radiation-Dominated Disks}
\shortauthors{Turner}
\slugcomment{Accepted by Ap.\ J.\ Letters}

\begin{document}
\title{On the Vertical Structure of\\ Radiation-Dominated Accretion
Disks}

\author{N. J. Turner\altaffilmark{1}}
\affil{Physics Department, University of California,
Santa Barbara CA 93106, USA}
\altaffiltext{1}{neal.turner@jpl.nasa.gov}

\begin{abstract}
The vertical structure of black hole accretion disks in which
radiation dominates the total pressure is investigated using a
three-dimensional radiation-MHD calculation.  The domain is a small
patch of disk centered 100 Schwarzschild radii from a black hole of
$10^8$ M$_\odot$, and the stratified shearing-box approximation is
used.  Magneto-rotational instability converts gravitational energy to
turbulent magnetic and kinetic energy.  The gas is heated by magnetic
dissipation and by radiation damping of the turbulence, and cooled by
diffusion and advection of radiation through the vertical boundaries.
The resulting structure differs in several fundamental ways from the
standard Shakura-Sunyaev picture.  The disk consists of three layers.
At the midplane, the density is large, and the magnetic pressure and
total accretion stress are less than the gas pressure.  In
lower-density surface layers that are optically thick, the magnetic
pressure and stress are greater than the gas pressure but less than
the radiation pressure.  Horizontal density variations in the surface
layers exceed an order of magnitude.  Magnetic fields in the regions
of greatest stress are buoyant, and dissipate as they rise, so the
heating rate declines more slowly with height than the stress.  Much
of the dissipation occurs at low column depth, and the interior is
cooler and less radiation-dominated than in the Shakura-Sunyaev model
with the same surface mass density and flux.  The mean structure is
convectively stable.
\end{abstract}


\section{INTRODUCTION}

Black holes in X-ray binary star systems and active galactic nuclei
accrete material through surrounding disks of gas.  When the accretion
rate is greater than about 1\% of the Eddington value, the radiation
pressure in the central regions of the disk may exceed the gas
pressure \citep{ss73}.  The disk structure and the radiation emerging
are governed by removal of angular momentum from the gas, and
conversion of released gravitational energy to photons.  In the
standard model, the angular momentum transfer and dissipation
processes are unspecified.  Both are assumed to occur at rates
proportional to the height-integrated total pressure, and the heating
is balanced by diffusion of photons to the disk surfaces \citep{ss73}.
The model is unstable to perturbations in mass accretion rate
\citep{le74} and heating rate \citep{ss76}, and to vertical convective
displacements \citep{bkb77}.  Any one of these instabilities might
substantially modify the disk structure.  However, if the stresses are
magnetic in nature, buoyancy of the fields may lead to stresses
scaling with the gas pressure alone, rather than the total pressure
\citep{sc89}.  In this case the viscous and thermal instabilities are
absent.  The instabilities may also be eliminated if part of the
accretion energy is dissipated outside the disk, in a hot corona
\citep{sz94}.  Observed properties that are not understood using the
standard Shakura-Sunyaev model include the X-ray emission \citep{em78}
and UV-optical spectra \citep{b04} of active galactic nuclei, and the
steady disk luminosities of some stellar-mass black hole X-ray binary
systems accreting at 1\% to 50\% of the Eddington rate \citep{gd04}.

The most likely physical mechanism of angular momentum transfer in
radiation-dom\-in\-at\-ed disks is magnetic stresses.  The
magneto-rotational instability or MRI \citep{bh91} leads to turbulence
in which the energy of differential orbital motion is converted to
magnetic fields and gas motions \citep{hgb96}.  In isothermal MHD
simulations in which heating and cooling are assumed to balance,
magnetic fields generated in the turbulence are partly expelled to
form magnetized coronae above and below the disk \citep{ms00}.

Disk structure may be affected also by the locations and rates of
heating and cooling.  Candidate physical heating mechanisms include
resistive dissipation of magnetic fields, microscopic viscous
dissipation of gas motions, and radiative damping of compressible
turbulence \citep{ak98}.  Cooling processes may include radiation
diffusion, convection \citep{ak01}, and photon bubble instability
\citep{g98}.  Here I examine the effects of heating and cooling on the
vertical structure of radiation-dominated disks, using a radiation-MHD
simulation.

\section{DOMAIN AND METHODS
\label{sec:method}}

The domain is a small patch of disk, centered 100~Schwarzschild radii
$R_S=2GM/c^2$ out from a black hole of $M = 10^8$ M$_\odot$, and
co-rotating at the central Keplerian orbital frequency $\Omega_0$.
The local shearing-box approximation is used, and Cartesian
coordinates $(x, y, z)$ correspond to distance from the origin along
the radial, orbital, and vertical directions, respectively
\citep{hgb95}.  The domain extends $1.5 R_S$ along the radial
direction, $6 R_S$ along the orbit, and $6 R_S$ either side of the
midplane, and is divided into $32\times 64\times 256$ zones.

The equations solved and numerical method differ from \cite{ts03} in
including the vertical component of the gravity of the black hole by a
term $-\rho\Omega^2_0 z$ in the equation of motion.  No viscosity of
the Shakura-Sunyaev type is used.  The frequency-averaged equations of
radiation MHD \citep{mm84,smn92} are integrated in the flux-limited
diffusion (FLD) approximation \citep{lp81}, using the Zeus code
\citep{sn92a,sn92b} with its FLD module \citep{ts01}.  Opacities due
to electron scattering and free-free processes are included.  The
equations are closed with a $\gamma=5/3$ ideal-gas equation of state.

Several dissipation mechanisms may act.  A total energy scheme is used
during the magnetic field update, so that numerical losses of magnetic
field are captured as gas heat.  The gas internal energy may also
increase through shock compression and associated artificial
viscosity.  The radiation energy increases when photons diffuse from
compressed regions, irreversibly extracting part of the work done in
compression \citep{ak98,ts02}.  Gas and radiation remain near mutual
thermal equilibrium while heating, owing to free-free absorption and
emission.  For a steady state, the total dissipation must be balanced
by diffusion of radiation, and advection of gas, radiation, and
magnetic energy through the vertical boundaries.

The azimuthal boundaries are periodic, the radial boundaries
shearing-periodic, and the vertical boundaries allow outflow but no
inflow.  The radiation flux into each vertical boundary zone is set
equal to that between the two adjacent active zones, except that the
flux is fixed at zero if needed to prevent radiation energy from
entering the domain.  Updated boundary fluxes are estimated using
photon diffusion coefficients from the previous timestep.  A density
floor is applied throughout.  In zones where the density falls below
0.2\% of the initial midplane value, mass is added to bring the
density up to the floor level.  The corresponding minimum optical
depth per zone in the vertical direction is 24.

\section{INITIAL STATE
\label{sec:ic}}

The surface mass density in the calculation sets the total optical
depth, and varies from its initial value only through outflows and the
density floor.  The net vertical magnetic flux affects the accretion
stress \citep{hgb95}, and is time-constant in the shearing-box
approximation.  Other aspects of the initial condition likely have
little effect on the outcome, since the structure is to be determined
by accretion stresses, dissipation, and cooling.

The initial condition is a \cite{ss73} model accreting at 10\% of the
Eddington value for luminous efficiency $0.1$.  In constructing the
initial state only, the ratio $\alpha$ of stress to total pressure is
set to $0.01$.  The resulting surface density is $1.1\times 10^6$
g~cm$^{-2}$.  The half-thickness $H$ of the Shakura-Sunyaev model is
$\frac{3}{4} R_S$.  The domain outside the model is filled with an
ambient medium of the floor density.  Since gravity increases with
height, while radiation flux is independent of height outside the
Shakura-Sunyaev model, the ambient medium is out of hydrostatic
balance and falls towards the midplane when the calculation starts.

The magnetic field is given zero net vertical flux, so that outflows
can completely demagnetize the domain.  The starting configuration is
an azimuthal flux tube of circular cross-section, with radius $0.75
H$.  The axis is offset from domain center by $+0.1 H$ in $x$ and
$+0.1 H$ in $z$.  Field strength in the tube is uniform at 2660 Gauss,
corresponding to 4\% of initial midplane gas plus radiation pressure.
The tube is twisted about its axis, giving maximum poloidal component
661 Gauss.  The maximum vertical MRI wavelength of 8 grid zones is
adequately resolved.  The domain-mean radial field is zero, and the
mean azimuthal field of 159 Gauss is much less than values that
develop later.  The calculation is begun with a small random poloidal
velocity in each grid zone.  The maximum amplitude of each velocity
component is 1\% of the midplane radiation sound speed.

\section{RESULTS}

During the first two orbital periods, the flux tube is stretched
radially by MRI, and its upper parts are lifted to the surface of the
Shakura-Sunyaev model by magnetic buoyancy.  By four orbits, the tube
is torn apart by MRI and spread throughout the region within $R_S$ of
the midplane.  Gravitational energy is released faster than assumed in
the initial Shakura-Sunyaev model, and the magnetized region expands
into the ambient medium.  After 13 orbits, magnetic fields are found
throughout the domain.

\subsection{Three Layers}

\begin{figure}
\epsscale{.9}
\plotone{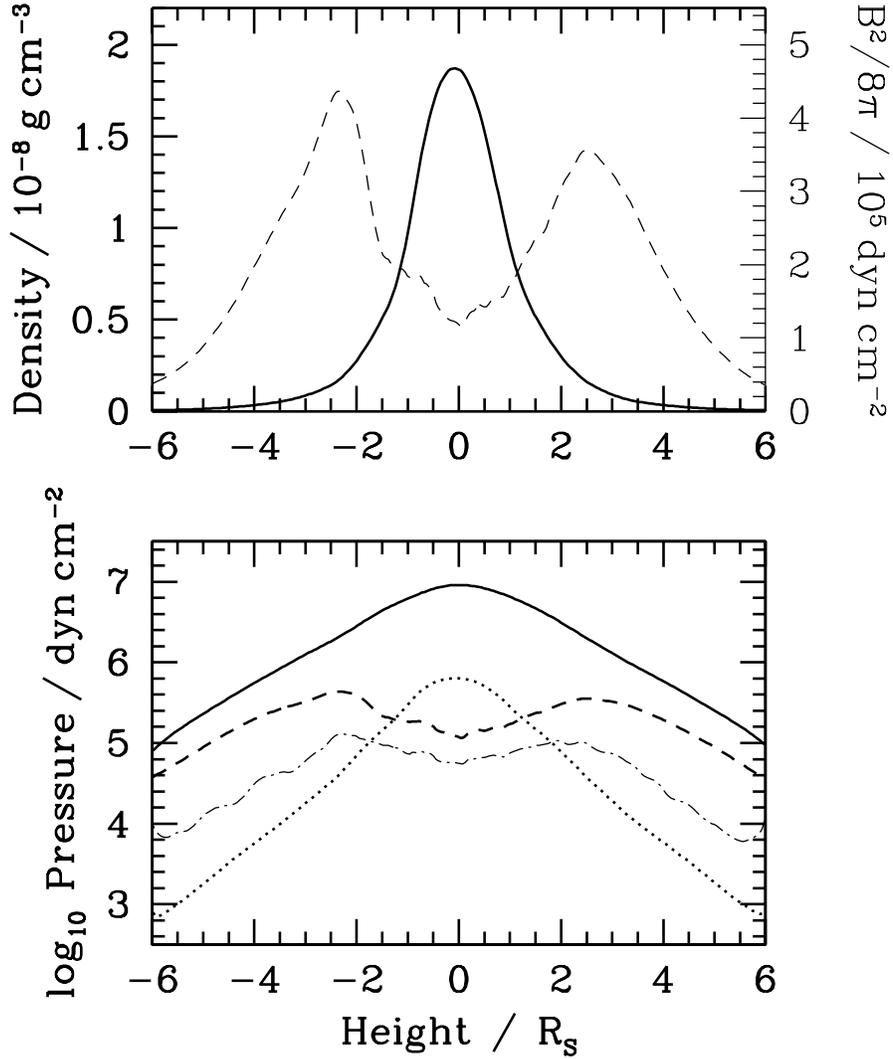}
\caption{Simulation results averaged horizontally and over time from
15 to 45 orbits.  The three layers may be seen at top.  Curves show
density (solid, left scale) and magnetic pressure (dashed, right
scale) versus height.  The surface layers are optically thick.  The
optical depth between each magnetic pressure peak and the nearby
boundary is $1.5\times 10^4$.  At bottom are plotted the pressures due
to radiation (solid), gas (dotted) and magnetic fields (dashed), and
the total accretion stress (dot-dashed).
\label{fig:3layer}}
\end{figure}

The horizontally-averaged structure present from 13 orbits on may be
divided into three layers (Figure~\ref{fig:3layer}).  In a dense layer
within about $1.5 R_S$ of the midplane, mean magnetic pressure and
total accretion stress are less than gas pressure.  In lower-density
surface layers, magnetic pressure and accretion stress are greater
than gas pressure, but less than the sum of gas and radiation
pressures.  The average structure is marginally convectively stable in
the midplane layer.  In the surface layers, the gas and radiation
Brunt-V\"ais\"al\"a frequencies are about equal to the orbital
frequency $\Omega_0$, and are real, indicating hydrodynamic convective
stability.

\begin{figure}
\epsscale{1.0}
\plotone{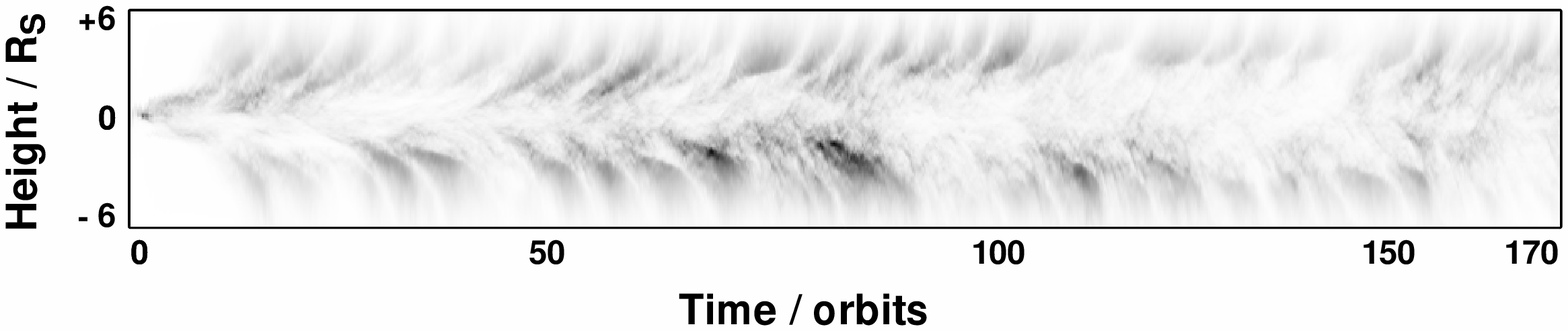}
\caption{Horizontally-averaged magnetic pressure versus height and
time.  The grey scale is linear from zero (white) to $3\times 10^6$
dyn cm$^{-2}$ (black).
\label{fig:tzplot}}
\end{figure}

Magnetic energy is produced fastest near the heights of greatest
accretion stress.  The fields are buoyant and rise towards the
boundaries (Figure~\ref{fig:tzplot}).  The rise speed is approximately
the Alfv\'en speed, as expected for magnetized regions exchanging heat
rapidly with their surroundings \citep{p75}.  The fields dissipate
numerically while rising, and the mean flux of magnetic energy through
the boundaries between 15 and 45 orbits is 6\% of the mean radiative
flux.  The horizontally-averaged magnetic pressure is dominated by the
azimuthal component throughout, and in most cases the sign of the
azimuthal field alternates between consecutive rising regions.  The
regular pattern seen in figure~\ref{fig:tzplot}, repeating about every
7 orbits, may result in part from the limited domain size.  Rising
regions in most cases fill the horizontal extent of the box.  In real
disks, adjacent buoyant regions may interact.

The photons may partly decouple from the turbulence, in
radiation-supported disks accreting via MRI, as the distance radiation
diffuses per orbit is about equal to the vertical MRI wavelength
\citep{ts03}.  Large density variations accompany loss of radiation
pressure support if magnetic pressure exceeds gas pressure
\citep{ts02}.  The results of the present calculation are consistent
with this picture.  Density variations on horizontal $x-y$ planes
exceed an order of magnitude near the upper and lower boundaries,
where the radiation diffusion distance is half the RMS vertical MRI
wavelength, and the mean magnetic pressure is more than 30 times the
gas pressure.  By contrast, horizontal density variations within $R_S$
of the midplane are only a factor two.  The average diffusion distance
is equal to the RMS vertical MRI wavelength, but the mean magnetic
pressure is less than the gas pressure, so magnetic forces squeeze the
gas less.  At the height $3.3 R_S$ where vertical magnetic pressure
peaks, the diffusion length is 0.3 times the vertical MRI scale.
Although radiation is better-coupled to gas motions, the magnetic
pressure is greater than the gas pressure.  Density variations are
intermediate between those at midplane and boundaries.

The vertical variation in magnetic field strength may be related also
to the diffusion of photons through the turbulence.  Linear MRI grows
slowly if magnetic pressure exceeds the pressure resisting compression
\citep{bb94,bs01}.  Near the midplane in the simulation, where
fluctuations are due largely to the competition of magnetic with gas
pressure forces, the turbulence saturates with magnetic pressure less
than gas pressure.  Around $3.3 R_S$, where squeezing is resisted by
both radiation and gas pressures, the magnetic pressure is greater
than the gas pressure.  Near the vertical boundaries, the rate of
arrival of magnetic fields from the interior exceeds the local rate of
field generation.

The vertical magnetic pressure profile is similar to those in
isothermal MHD simulations of gas-supported disks by \cite{ms00}.
However the magnetized surface layers here are optically thick, and
lie inside the disk photosphere.  Horizontally-averaged magnetic
pressure is everywhere less than gas plus radiation pressure.  The
total accretion stress at the midplane is half the maximum in the
surface layers, whereas in the \cite{ms00} fiducial calculation, total
stress in the gas-dominated region varies irregularly with height.

\subsection{Comparison with Shakura-Sunyaev Model}

\begin{figure}
\epsscale{.8}
\plotone{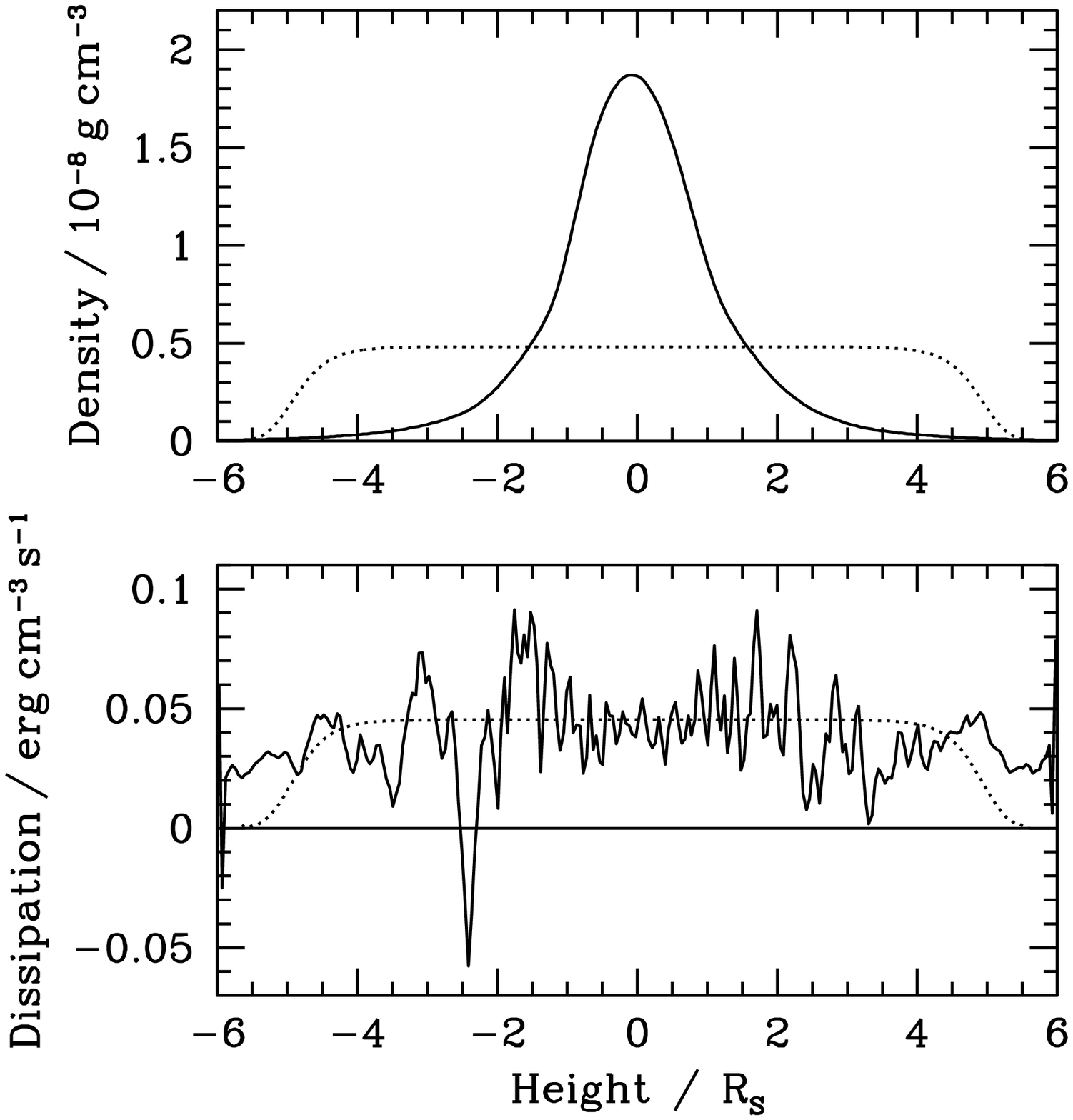}
\caption{Simulation results (solid) compared with the Shakura-Sunyaev
model (dotted) having the same surface density and radiation flux.
Density is shown at top, total dissipation below.  The simulation
results are averaged horizontally, and over time from 15 to 45 orbits.
\label{fig:alpha}}
\end{figure}

The averaged simulation results may be compared against the
Shakura-Sunyaev model with the same surface density and radiation
flux.  The model has $\alpha=0.0013$ and accretion rate 64\% of the
Eddington value.  Density is almost independent of height in the model
interior, while in the simulation the density is
centrally-concentrated (Figure~\ref{fig:alpha}).  In the
Shakura-Sunyaev picture, it is often assumed that dissipation is
locally proportional to density.  By contrast, much of the dissipation
in the simulation occurs in regions of low density.  Photons escape
easily from the surface layers, so the interior is cooler than in the
Shakura-Sunyaev model.  The midplane ratio of radiation to gas
pressure is 160 in the model, and 14 in the simulation.  The thermal
time in the Shakura-Sunyaev model is 370 orbits, while in the
simulation the cooling time, or ratio of total energy content to
cooling rate, is just 19 orbits.  Dissipation falls off more slowly
with height than the stress, as fields tend to rise between generation
and dissipation.

\subsection{Thermal Balance}

The radiation content of the flow is maintained through accretion.
Between 15 and 45 orbits, the energy released is 157\% of the mean
domain-integrated total energy, while the total energy is almost
constant, increasing just 2\%.  The source of power, the differential
rotation, is transformed into turbulent magnetic and kinetic energy by
magnetic stresses.  The turbulence is converted to heat mainly through
numerical losses of magnetic fields.  Radiation damping contributes
29\% of the heating within $R_S$ of the midplane, where it is readily
distinguished from expansion associated with magnetic buoyancy.
Artificial viscous heating, mostly in shocks, makes up 12\% of the
overall dissipation.  The dissipation is approximately balanced by
cooling.  The main cooling processes are diffusion and advection of
radiation through the vertical boundaries, with diffusion carrying
two-thirds of the energy flux.  Mass lost through the boundaries is
balanced by mass added to maintain densities above the floor, and
total mass increases by less than 1\% from 15 to 45 orbits.  The
energy flux is 73\% of the energy input, and the remainder disappears
partly through numerical losses of kinetic energy.  A complete
energy-conserving scheme may be useful for future calculations.

Starting at 50 orbits, the disk heats, then cools.  Total energy
increases roughly linearly by a factor 2.5 up to 90 orbits, decreases
linearly until 145 orbits, then is again steady.  The run ends at 170
orbits, or 8 simulation thermal times after first saturation.  The
heating rate varies, while the diffusion cooling rate is approximately
time-constant.  No exponentially-growing thermal instability is seen.
However, during the hot period, the disk expands, and losses through
the vertical boundaries reduce the mass by almost half.  The absence
of runaway heating may be due to the mass loss, rather than internal
thermal stability.

\section{CONCLUSIONS}

A patch of radiation-dominated accretion disk lying 100 Schwarzschild
radii from a $10^8$ M$_\odot$ black hole is simulated including
physical processes of energy release, dissipation, and cooling.  The
patch has surface density $10^6$ g~cm$^{-2}$, and its net vertical
magnetic flux is zero.  The structure that develops has density
greatest at the midplane, stresses greatest in optically-thick surface
layers, and dissipation found throughout but intermittent in time and
space.  The stresses result from magnetic forces.  The fields in the
surface layers are buoyant.  Heating occurs mainly by numerical
dissipation of the fields, and physical radiation damping of the
turbulence, while cooling occurs by diffusion and advection of
radiation through the boundaries.  The vertical structure may prove to
vary with mass of the central body, location in the disk, surface
density, and magnetic flux.

The time-averaged structure in the simulation is hydrodynamically
convectively stable.  Thermal instability may be prevented by outflow
through the boundaries.  Viscous stability is not tested, as there is
no net radial flow in the shearing box.  Photon bubble instability may
be present.  The fastest linear modes \citep{bs03} in the
time-averaged structure are found in the surface layers, have
wavelengths shorter than the vertical grid spacing, and grow at
$0.9\Omega_0$.  Resolved modes, with wavelengths greater than six grid
zones, are expected to grow more slowly than the MRI.  No clear signs
of these modes are found.  The likely direct effect of photon bubbles
is to further reduce the cooling time.

\begin{acknowledgments}
The methods used here were developed with Jim Stone and Takayoshi
Sano.  I benefited also from discussions with Julian Krolik and Omer
Blaes.  The work was supported by the U. S. National Science
Foundation under grant AST-0307657.
\end{acknowledgments}


\end{document}